\RequirePackage{amsmath}
\documentclass[12pt]{iopart}
\usepackage{graphicx,psfrag}
\usepackage{iopams}
\usepackage{epstopdf}
\begin{document}
\title[Rashba-induced spin electromagnetic fields]
{
Rashba-induced spin electromagnetic fields in a strong sd coupling regime
}
% monopole in spin pumping system (clean limit)
\author{ Noriyuki Nakabayashi$^1$ and Gen Tatara$^2$}
\address{%\affiliation{
$^1$ Graduate School of Science and Engineering, Tokyo Metropolitan University, Hachioji, Tokyo 192-0397 Japan \\
$^2$  RIKEN Center for Emergent Matter Science (CEMS), 
2-1 Hirosawa, Wako, Saitama, 351-0198 Japan 
}

\begin{abstract}
Spin electromagnetic fields driven by the Rashba spin-orbit interaction, or Rashba-induced spin Berry's phase,  in ferromagnetic metals is theoretically studied based on the Keldysh Green's function method. 
Considering a limit of strong sd coupling without spin relaxation (adiabatic limit),  
the spin electric and magnetic fields are determined by calculating transport properties.
The spin electromagnetic fields turn out to be expressed in terms of a Rashba-induced effective vector potential, and thus they satisfy the Maxwell's equation.
In contrast to the conventional spin Berry's phase, the Rashba-induced one is linear in the gradient of magnetization profile, and thus can be extremely large even for slowly varying structures.
We show  that the Rashba-induced spin Berry's phase exerts the Lorentz force on spin resulting in a giant spin Hall effect in magnetic thin films in the presence of magnetization structures. 
Rashba-induced spin magnetic field would be useful to distinguish between topologically equivalent magnetic structures. 
\end{abstract}

\date{\today}                                           % Activate to display a given date or no date
%\date{\hfill}       %no date

% bold
\newcommand{\vect}[1]{\boldsymbol{#1}}
\newcommand{\bl}[1]{\boldsymbol{#1}}
\newcommand{\kv}{\boldsymbol{k}}
\newcommand{\qv}{\boldsymbol{q}}
\newcommand{\pv}{\boldsymbol{p}}
\newcommand{\xv}{\boldsymbol{x}}
\newcommand{\nv}{\boldsymbol{n}}
\newcommand{\vv}{\boldsymbol{v}}
\newcommand{\Ev}{\boldsymbol{E}}
\newcommand{\Bv}{\boldsymbol{B}}
\newcommand{\Riv}{\boldsymbol{R}_{\rm i}}
\newcommand{\Pv}{\boldsymbol{P}}
\newcommand{\Fv}{\boldsymbol{F}}
\newcommand{\jv}{\boldsymbol{j}}
\newcommand{\rv}{\boldsymbol{r}}
\newcommand{\Av}{\boldsymbol{A}}
\newcommand{\Mv}{\boldsymbol{M}}
\newcommand{\mv}{\boldsymbol{m}}
\newcommand{\alphav}{\boldsymbol{\alpha}}
\newcommand{\sigmav}{\boldsymbol{\sigma}}
\newcommand{\nablav}{\boldsymbol{\nabla}}
% \lambda_{\rm R} etc.
\newcommand{\lamR}{\lambda_{\rm R}}
\newcommand{\ER}[1]{E_{{\rm R},{#1}}}
\newcommand{\bER}{\boldsymbol{E}_{\rm R}}
\newcommand{\lami}{\lambda_{\rm i}}
% \lambda_{\textup{so}}
\newcommand{\lamso}{\lambda_{\textup{so}}}
% v_{\textup{so}}
\newcommand{\vso}{v_{\textup{so}}}
% n_{\textup{so}}
\newcommand{\nso}{n_{\textup{so}}}
% imp
\newcommand{\nimp}{n_{\rm i}}
\newcommand{\vimp}{v_{\rm i}}
%H
\newcommand{\Hso}{H_{\rm so}}
\newcommand{\Hem}{H_{\rm em}}
\newcommand{\Himp}{H_{\rm imp}}
%Nimp
\newcommand{\Nimp}{N_{\rm imp}}
%c^{\dag}
\newcommand{\cdag}{c^{\dag}}
\newcommand{\adag}{a^{\dag}}
% j_s
\newcommand{\js}[1][]{j_{{\rm s}#1}}
\newcommand{\jsv}{\boldsymbol{j}_{\rm s}}
%\overset{\leftrightarrow}{\nabla}
\newcommand{\nablalr}{\overset{\leftrightarrow}{\nabla}}
\newcommand{\bnablalr}{\overset{\leftrightarrow}{\bl{\nabla}}}
% \mathcal{T}
%\newcommand{\Tau}{\mathcal{T}}
%\newcommand{\Tauso}[1]{\mathcal{T}_{\vso}^{#1}}
%\newcommand{\TausoA}[1]{\mathcal{T}_{\vso}^{A,#1}}
% average
\def\average#1{\left\langle {#1} \right\rangle}
% tr
%\newcommand{\tr}{{\rm tr}}
% \rm
\newcommand{\so}{{\rm so}}
\newcommand{\s}{{\rm s}}
\newcommand{\R}{{\rm R}}
\newcommand{\B}{{\rm B}}
\newcommand{\M}{{\rm M}}
\newcommand{\eff}{{\rm eff}}
% \nonumber
\newcommand{\nn}{\nonumber}
% g^{r}, g^{a}
\newcommand{\gr}{g^{r}}
\newcommand{\ga}{g^{a}}

\maketitle
%\vspace*{-2cm}
%\section*{}
%\subsection{}

\section{Introduction}
\newcommand{\Asv}{\Av_{\rm s}}  %\newcommand{\Bsv}{{\bm B}_{\rm s}}
\newcommand{\Bsv}{\Bv_{\rm s}}  %\newcommand{\Bsv}{{\bm B}_{\rm s}}
\newcommand{\Esv}{\Ev_{\rm s}}  %\newcommand{\Esv}{{\bm E}_{\rm s}}
\newcommand{\lt}{\left}
\newcommand{\rt}{\right}
\newcommand{\kf}{k_{\rm F}}
\newcommand{\ef}{\epsilon_{\rm F}}
\newcommand{\sd}{{\rm sd}}
\newcommand{\Sv}{\boldsymbol{S}}

Electromagnetism is one of the most important physical phenomena which our modern technologies are based on. 
Precise design of electric devices is possible owing to the Maxwell's equation, which describes a mathematical structure of the electromagnetism.
Such a structure is not unique to the electromagnetism in the vacuum; it rather arises whenever there is a U(1) gauge symmetry.
In solids, several possibilities of U(1) gauge symmetry are known to emerge.
A simple example is a metallic magnet. 
A magnet is an ensemble of many spins, which are governed quantum mechanically by a non-commutative algebra of SU(2).
When a magnet has a non uniform magnetization texture, a motion of conduction electrons is equivalent to that of a particle in a curved space; 
To put mathematically, it is described by a gauge field having SU(2) gauge symmetry.
When an interaction between the magnetization and conduction electron (sd interaction) is strong, as is the case in most metallic ferromagnets, symmetry breaking of spin space occurs, and only one component of the SU(2) gauge field survives, resulting in an emergent U(1) gauge field or emergent spin electromagnetism \cite{Volovik87}.

\begin{figure}[hbtp]
\begin{center}
\includegraphics[width=0.4\hsize]{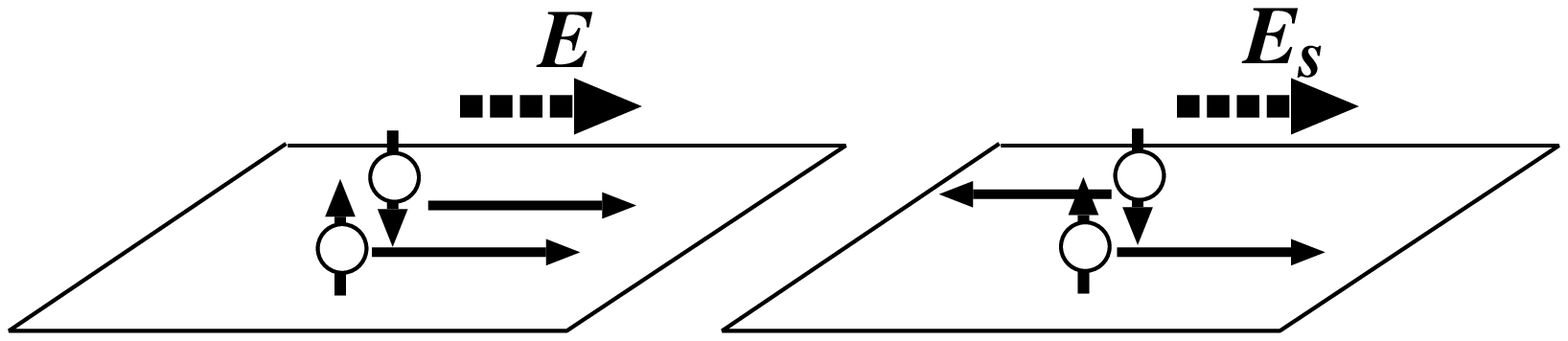}
\includegraphics[width=0.4\hsize]{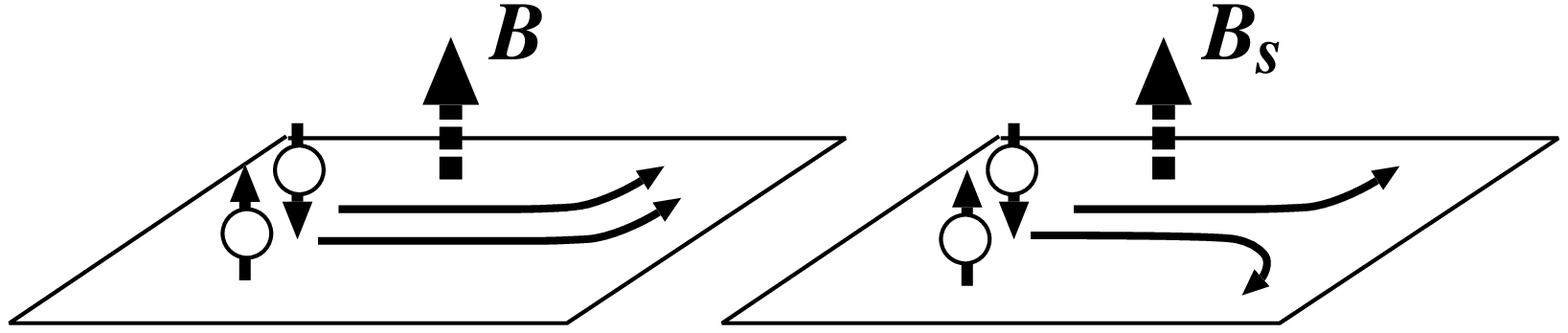}
 \caption{
Schematic figure showing roles of conventional electromagnetic fields, $\Ev$ and $\Bv$, and spin electromagnetic fields, $\Esv$ and $\Bsv$.
While $\Ev$ and $\Bv$ drives conduction electrons having two different spins in the same direction, $\Esv$ and $\Bsv$ drives the two spins in the opposite direction, inducing spin current and spin Hall effect, respectively.
\label{fig:EandEs}}
\end{center}
\end{figure}

By definition, spin electric field drives spin up and down conduction electrons in opposite directions, inducing a spin current, $\js$ (Fig. \ref{fig:EandEs}). 
In ferromagnetic metals, driven spin current is usually associated with a charge current, given by $j=\js/P$, where $P$ is a parameter representing spin polarization of carrier.
Therefore,  the spin electric field or spin motive force can be detected as a voltage, as was observed in the case of motion of a domain wall, vortex and skyrmions \cite{Yang09,Tanabe12,Schulz12}.
It is notable that a voltage generation by a moving domain wall was predicted by Berger \cite{Berger86}  based on a phenomenological argument earlier than a gauge field argument by Volovik. 
In the same manner as spin electric field, the spin magnetic field, conventionally called the spin Berry's phase, exerts the spin-dependent Lorentz force on the conduction electrons, inducing spin Hall effect, and the spin Hall effect is detected as an anomalous Hall effect \cite{Nagaosa10}.
%The emergent field, which we call spin electromagnetic field, is coupled to the electrons' spin and is thus useful in controlling spin transports, i.e., in spintronics.
To understand the whole structure of the emergent spin electromagnetic field is of essential importance in spintronics.
%%%%%%%%%%%%%%%%%%%%%%%%
%%%%%%%%%%%%%%%%%%%%%%%%%%%%%%%%%%%%%%%%%%%%%%%%%%%%%%%

When spin-orbit interaction is present, SU(2) gauge symmetry is affected resulting in novel contributions to spin electromagnetic fields. 
The spin-orbit correction to the spin electric field has been theoretically studied in \cite{Duine09,Kim12,Tatara12}.
Of particular current interest is the effect of Rashba spin-orbit interaction arising from the breaking of inversion symmetry at surfaces and interfaces.
Takeuchi et al. investigated weak sd coupling regime and found a spin electric field proportional to 
$\alphav_{\rm R}\times(\nv\times\dot{\nv})$\cite{Takeuchi12,Tatara12}, where $\alphav_{\rm R}$ is the Rashba electric field and ${\nv}$ is a unit vector representing the local magnetization  direction,
while 
Kim et al. obtained a different form of 
$\alphav_{\rm R}\times\dot{\nv}$ in a strong sd coupling regime \cite{Kim12}.
It was shown later in \cite{Tatara_smf13} that the contribution found in \cite{Takeuchi12} arises in the strong sd limit when spin relaxation is included.
The two contributions thus coexist in general.
A unique feature of the Rashba-induced spin electric field is that it arises even from spatially uniform magnetization, in contrast to the conventional one induced by inhomogeneous magnetization textures.

For deriving an expression for spin electromagnetic fields, a gauge field argument based on a U(1) gauge invariance is useful \cite{Volovik87} if in the absence of spin relaxation.
It can be also calculated directly by evaluating a force acting on spin, which is proportional to the time-derivative of the current density \cite{Kim12,Tatara_smf13}.
Spin electromagnetic field is also accessible by a transport calculation. 
In fact, in \cite{Takeuchi12}, the fields were identified by calculating the induced electric current density and then comparing the result with a general expression resulting from the Maxwell's equation, $\jv=\sigma_{\rm s} \Esv+\nablav\times \Bsv -D_{\rm s}\nablav\rho_{\rm s}$, where $\sigma_{\rm s}$ is spin conductivity, $D_{\rm s}$ is diffusion constant for spin and $\rho_{\rm s}$ is spin density.
This approach is highly useful to study the weak coupling regime, where adiabatic component of spin gauge field can not be defined.
As noted in \cite{Takeuchi12,Tatara12}, the information in the pumped current, however, is not enough to completely determine the two fields and  effective permittivity and permeability, and additional information is necessary.
Besides, it is not clear whether all of the contributions to the current expressed as rotation of certain vectors are to be interpreted always as due to an effective magnetic field.
This issue is answered by investigating the Hall effect;
If the contribution is really induced by an effective magnetic field, the field should exert the Lorentz force on the electron spin and induce spin Hall effect.
The aim of this paper is to determine a spin magnetic field uniquely by calculating the spin Hall effect, and explore structure of the spin electromagnetism induced by the Rashba interaction and dynamic magnetization structures.
We shall show that the whole contribution pumped current that is written as a rotation of a certain vector is indeed to be identified as an effective magnetic field, at least in the present system of strong sd coupling limit with the Rashba interaction.

The spin electromagnetic fields studied here are generalized spin Berry's phase including the Rashba effects. 
In contrast to the conventional spin Berry's phase, the Rashba-induced one is linear in the gradient of magnetization profile, and thus it would dominate in the case of slowly varying magnetization.
We show that the spin electric and magnetic fields are estimated to be extremely strong like 2.5 kV/m and 2.5 kT, respectively, for a strong Rashba interaction induced at surfaces \cite{Ast07} when the frequency and the length scale of magnetization profile are 1GHz and 1nm, respectively.

We demonstrate that the Rashba-induced effective spin electromagnetic fields in the strong sd coupling limit without spin relaxation are described totally by an effective U(1) gauge field, $\Av_{\rm R} = \frac{m}{e\hbar}(\alphav_{\R}\times\nv)$ ($m$ and $(-e)$ are the electron mass and charge, respectively), 
 at the linear order in the Rashba interaction.
There is therefore no monopole in the present system, in contrast to what was observed in \cite{Takeuchi12} in the weak sd coupling regime.
In the light of the present result and that of \cite{Takeuchi12}, spin relaxation seems to be essential for an emergence of monopole, as claimed in \cite{Takeuchi12}.
In fact, as will show below, the spin electric field in the absence of spin relaxation is proportional to 
$\Esv\propto \alphav_{\rm R}\times\dot{\nv}$, where $\alphav_{\rm R}$ 
is the Rashba-field, in agreement with the result of \cite{Kim12}.
Its rotation, $\nablav\times\Esv$, is thus written as a time derivative of $\nablav\times(\alphav_{\rm R}\times{\nv})$, and as we will show in the present paper, it is in fact written totally by a time-derivative of spin magnetic field, $\Bsv\propto \nablav\times(\alphav_{\rm R}\times{\nv})$, i.e., $\nablav\times\Esv=-\dot{\Bsv}$.
The two fields satisfy thus the conventional Faraday's law and there is no monopole term.
In contrast, the spin electric field found in \cite{Takeuchi12,Tatara_smf13} in the presence of spin relaxation is
$\Esv\propto \alphav_{\rm R}\times(\nv\times\dot{\nv})$, and cannot be written by a time derivative of any local quantity.
It therefore follows that $\nablav\times\Esv+\dot{\Bsv} \equiv -\jv_{\rm m}$ is non-vanishing for any local function $\Bsv$, resulting in a finite current of spin damping monopole \cite{Takeuchi12}, $\jv_{\rm m}$.
The spin damping monopole was argued to be essential in spin-charge conversion in dynamic magnetization structures such as in the inverse spin Hall effect \cite{Saitoh06}.

%\clearpage
%\vspace{-10pt}

%%%%%%%%%%%%%%%%%%%%%%%%%%%%%%%%%%%%%%%%%%%%%%%%%%%%%%%%%%%%%%%%%%%%%%%%%%%%
\section{Calculation of electric current}\label{sec:rashba_surrent}
In this section, we derive an expression for spin electromagnetic fields by calculating an electric current induced by magnetization dynamics and the Rashba interaction in a metallic ferromagnet. The Lagrangian of the system is 
\begin{align}
 {\cal L}
 = & \int d^3 r \cdag
\biggl[
	\rmi\hbar\partial_t
	+\left(\frac{\;\hbar^2}{2m}\nabla^2 +\epsilon_F \right)
	+\Delta_{\rm sd}\left(\bl{n}\cdot\bl{\sigma}\right)
	-\frac{\rmi}{2}\,\alphav_{\rm R}\cdot
	\Bigl( \bnablalr\times\bl{\sigma}\Bigr)
	-\vimp
\biggr]c,
\label{eq:Lnotgaugetrans}
\end{align}
where $c$ and $\cdag$ are annihilation and creation operators of
conduction electron respectively,
$\alphav_{\rm R}$ represents the strength and the direction of the Rashba interaction, 
$\nv$ is the unit vector parallel to local magnetization,
$\vimp$ is the random potential caused by spin-independent impurities,
$\bl{\sigma}$ is a Pauli matrix vector,  
and $\Delta_{\rm sd}$ is the strength of sd exchange interaction.
We consider the limit where $\Delta_{\rm sd}$ is large, and perform a local
spin gauge transformation (rotation in spin space) defined by
 $a\equiv Uc$, where $a$ is the annihilation operator in a rotated space
and $U\equiv \mv\cdot\sigmav,$ where
$\mv\equiv\left(\sin\frac{\theta}{2}\cos\phi,
\sin\frac{\theta}{2}\sin\phi, \cos\frac{\theta}{2}\right)$, ($\theta$ and
$\phi$ are polar coordinates of $\nv$), is the
rotation matrix in spin space \cite{TKS_PR08}. 
This transformation diagonalize the sd interaction.
%Equation 
(\ref{eq:Lnotgaugetrans}) then reads
\begin{align}
  {\cal L}
 = & \int d^3 r 
\biggl\{
   \rmi\hbar\adag\partial_t a -\hbar \adag A_{\s,t} a
   +\adag\left(\frac{\;\hbar^2}{2m}\nabla^2 +\epsilon_F \right)a
   +\rmi\frac{\;\hbar^2}{2m}A_{\s,j}^\ell \Bigl(\adag\nablalr_{\!\! j} 
   \sigma_\ell a \Bigr)
   -\frac{\;\hbar^2}{2m}\adag A_{\s}^2 a
   \nn \\ 
   & \qquad\qquad
   +\Delta_{\rm sd}\adag \sigma_z a
   -\frac{\rmi}{2}\alpha_{\R,j} \epsilon_{jk\ell}R_{\! \ell n}
   \Bigl[ \Bigl(\adag\sigma_n \nablalr_{\! k}a\Bigr) +2\rmi A_{\s,k}^n \adag a \Bigr]
   -\vimp\adag a
\biggr\},
\label{eq:Lgaugetrans}
\end{align}
where $A_{\s,\mu}\equiv
A_{\s,\mu}^\ell\sigma_\ell\equiv[\mv\times(\partial_\mu \mv)]_\ell 
\sigma_\ell,$ is the spin SU(2) gauge field and $R_{\ell n} \equiv 2m_\ell m_n
-\delta_{\ell n}$ is a rotation matrix element.
Summation is assumed for repeated indices 
($A_{\s,\mu}^\ell\sigma_\ell\equiv\sum_{\ell=x,y,z}A_{\s,\mu}^\ell\sigma_\ell$).

\begin{figure}[hbp]
%  \psfrag{R}{$\alpha_{\R}$}
\begin{center}
\includegraphics[width=130pt]{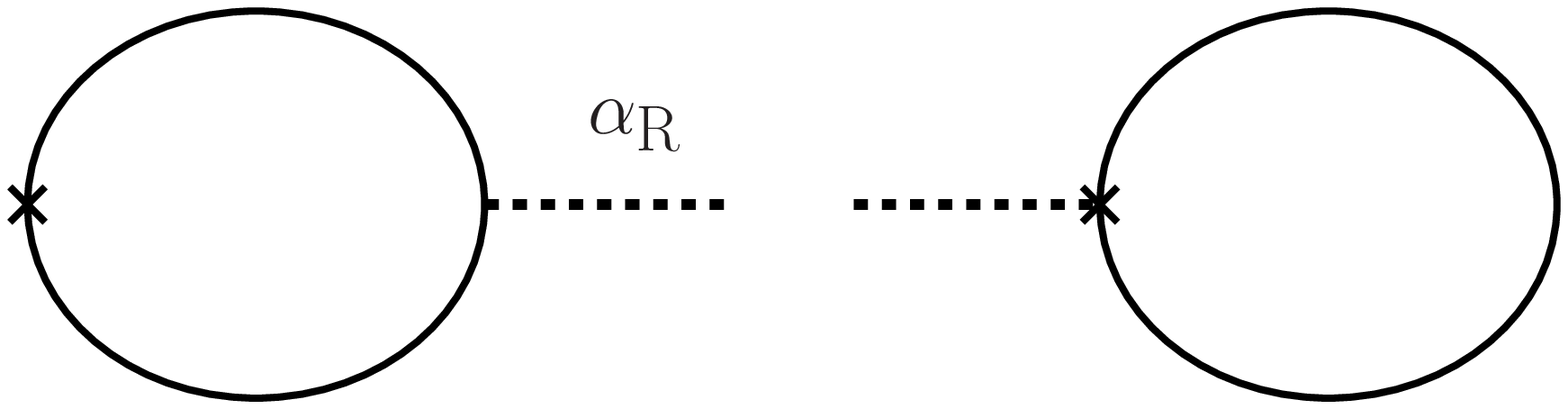}
\includegraphics[width=130pt]{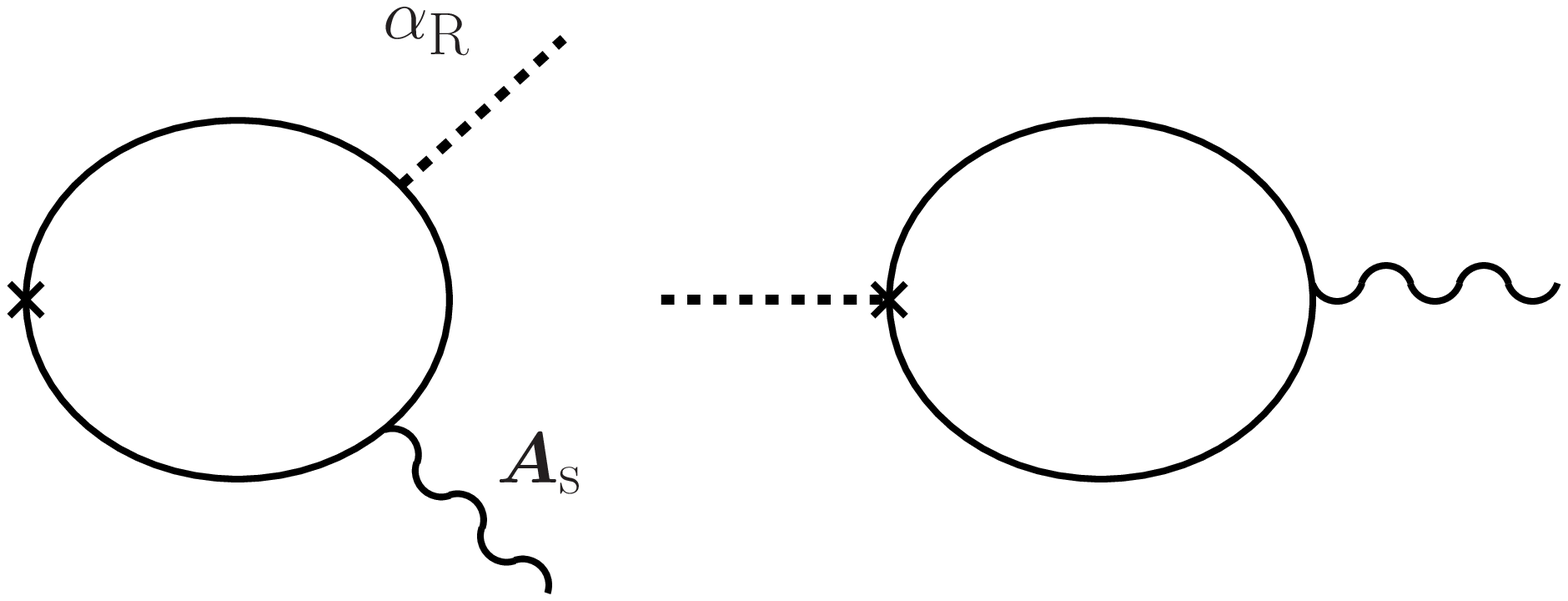}
\includegraphics[width=130pt]{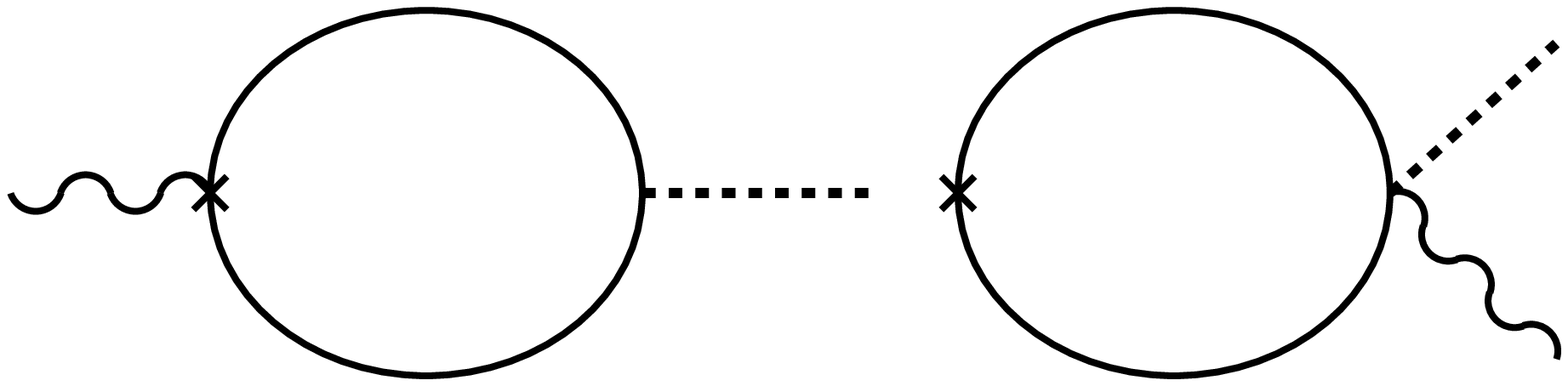}
  \caption{The Feynman diagrams representing the lowest order local contributions
     of electric current
     driven by the Rashba spin-orbit interaction, $\alpha_{\rm R}$, and spin gauge field, $A_{\rm S}$. A cross represents a current vertex.
     The dotted line and the wave line denote the
     Rashba interaction and the spin gauge interaction respectively.
\label{fig:jlocal_dia}}
\end{center}
\end{figure}

For estimating the effective field, we calculate the electric current
induced by the Rashba interaction following the approach of \cite{Takeuchi12}.
The electric current density written in terms of $a$ and $\adag$
 is %(we define the electric charge equal to $-e < 0$)
\begin{align}
 j_i =&
\frac{\rmi e\hbar}{2m}\average{\adag\nablalr_i a}
-\frac{e\hbar}{m}A_{\s,i}^\ell \average{\adag\sigma_\ell a}
-\frac{e}{\hbar}\epsilon_{ijk}\alpha_{\R,j}R_{k\ell}\average{\adag\sigma_\ell
 a},
\label{eq:jdef}
\end{align}
where $\average{\;}$ denotes the expectation value in the ground state.
%Since the effective field the case of no Rashba interaction are well
%known
We calculate (\ref{eq:jdef}) at the first order of Rashba interaction.
Generally, electric current pumped by dynamic spins is a sum of local terms and
diffusion terms \cite{Hosono10,Takeuchi10}, but we here look into the local terms only, which represents the local effect of the effective fields.
The leading contributions of local electric current density are 
diagramatically depicted in Fig.\ref{fig:jlocal_dia}.
The contribution represented by the first two diagrams in figure \ref{fig:jlocal_dia} reads
\begin{align}
 j^{12}_i(\rv,t) =&
-\frac{\rmi e\hbar^2}{m}\alpha_{\R,\ell}\epsilon_{\ell mn}
\sum_{\kv,\pv}\sum_{\omega,\bar{\Omega}}
\rme^{-\rmi\pv\cdot\rv+\rmi\bar{\Omega}t}
R_{no}(\pv,\bar{\Omega})
\tr\!\Bigl[
k_i k_m 
g_{\kv-\frac{\pv}{2},\omega-\frac{\bar{\Omega}}{2}}
\sigma_o g_{\kv+\frac{\pv}{2},\omega+\frac{\bar{\Omega}}{2}}
+
\delta_{im}\frac{m}{\;\hbar^2}
\sigma_o g_{\kv,\omega}
\Bigr]^<,
\label{eq:j0}
\end{align}
where $g_{\kv,\omega}$ is the counter ordered Green's function of conduntion electron 
with wave vector $\kv$ and angular frequency $\omega$, and $^<$
represents a lesser component.
It includes the elastic lifetime due to the impurities, $\tau$, and is a $2\times2$ matrix in spin space.
The contribution of remaining diagrams in figure \ref{fig:jlocal_dia} reads
\begin{align}
j^{3-6}_i =\,
& -\frac{\rmi e\hbar^2}{m}\alpha_{\R,\ell}\epsilon_{\ell mn}\!\!
\sum_{\kv,\qv,\pv}\sum_{\omega,\Omega,\bar{\Omega}}
\rme^{-\rmi (\qv+\pv)\cdot\rv+\rmi (\Omega+\bar{\Omega}) t}
R_{no}(\pv,\bar{\Omega})
\nn \\
&\times
\tr\!\Bigl[
k_i \Bigl(k+\frac{q}{2}\Bigr)_m \hbar 
J_{\mu}\!\Bigl(\kv-\frac{\pv}{2}\Bigr) 
A_{\s,\mu}^j (\qv,\Omega)
g_{\kv-\frac{\qv}{2}-\frac{\pv}{2},\omega-\frac{\Omega}{2}-\frac{\bar{\Omega}}{2}}
\sigma_j g_{\kv+\frac{\qv}{2}-\frac{\pv}{2},\omega+\frac{\Omega}{2}-\frac{\bar{\Omega}}{2}} \sigma_o g_{\kv+\frac{\qv}{2}+\frac{\pv}{2},\omega+\frac{\Omega}{2}+\frac{\bar{\Omega}}{2}} +c.c.
\nn \\
&\qquad+
k_i A_{\s,m}^o(\qv,\Omega)
g_{\kv-\frac{\qv}{2}-\frac{\pv}{2},\omega-\frac{\Omega}{2}-\frac{\bar{\Omega}}{2}} g_{\kv+\frac{\qv}{2}+\frac{\pv}{2},\omega+\frac{\Omega}{2}+\frac{\bar{\Omega}}{2}}
%\nn \\
%&\qquad
+A_{\s,i}^j (\qv,\Omega) k_m
\sigma_j g_{\kv-\frac{\pv}{2},\omega-\frac{\bar{\Omega}}{2}}
\sigma_o g_{\kv+\frac{\pv}{2},\omega+\frac{\bar{\Omega}}{2}}
\nn \\
&\qquad+
\delta_{im}\frac{m}{\;\hbar^2} \hbar J_{\mu}\!(\kv)A_{\s,\mu}^j(\qv,\Omega)
\sigma_o g_{\kv-\frac{\qv}{2},\omega-\frac{\Omega}{2}}
\sigma_j g_{\kv+\frac{\qv}{2},\omega+\frac{\Omega}{2}}
\Bigr]^<,
\label{eq:j1}
\end{align}
where $J_t(\kv) \equiv 1, J_i(\kv)\equiv\frac{\hbar}{m}k_i$.%,\,
%and $\jv^{\rm diff}$ is the diffusion contribution described the diagram
%in fig....(this contribution calculated later.)

%%%%%%%%%%%%%%%%%%%%%%%%%%%%%%%%%%%%%%%%%%%%%%%%%%%%%%%
%%%%%%%%%%%%%%%%%%%%%%%%%%%%%%%%%%%%%%%%%%%%%%%%%%%%%%%

First, we culculate (\ref{eq:j0}).
The lesser components in (\ref{eq:j0}) are given as
\begin{align}
(g_{\kv-\frac{\pv}{2},\omega-\frac{\bar{\Omega}}{2}}\sigma_o g_{\kv+\frac{\pv}{2},\omega+\frac{\bar{\Omega}}{2}})^<
=&\;
 f_{\omega-\frac{\bar{\Omega}}{2}}\bigl(\ga_{\kv-\frac{\pv}{2},\omega-\frac{\bar{\Omega}}{2}}-\ga_{\kv+\frac{\pv}{2},\omega+\frac{\bar{\Omega}}{2}}\bigr)\sigma_o
 \ga_{\kv+\frac{\pv}{2},\omega+\frac{\bar{\Omega}}{2}} \nn \\
&\quad
+f_{\omega+\frac{\bar{\Omega}}{2}}\gr_{\kv-\frac{\pv}{2},\omega-\frac{\bar{\Omega}}{2}}\sigma_0 \bigl(\ga_{\kv+\frac{\pv}{2},\omega+\frac{\bar{\Omega}}{2}}-\gr_{\kv+\frac{\pv}{2},\omega+\frac{\bar{\Omega}}{2}}\bigr)
 \\
g_{\kv,\omega}^< =&\; f_{\omega}(\ga_{\kv,\omega}-\gr_{\kv,\omega}),
\end{align}
where $\gr_{\kv,\omega} $ and $\ga_{\kv,\omega} $  are retarded and advaiced Green's function and $f_\omega=\frac{1}{\rme^{\hbar\beta\omega}+1}$ is the Fermi distribution
function ($\beta$ is the inverse temperature).
Therefore $j^{12}$ is calculated as
\begin{align}
j_i^{12} =&\;
\sum_{\pv,\bar{\Omega}}\rme^{-\rmi\pv\cdot\rv+i\bar{\Omega}t}
\xi_{ij}\Bigl(
\alphav_{\R}\times\nv(\pv,\bar{\Omega})\Bigr)_j
\label{eq:j0_xi}
\end{align}
where
\begin{align}
 \xi_{ij}&\equiv \frac{\rmi e\hbar^2}{m}\sum_{\kv,\omega}\sum_{\sigma=\pm} \sigma
\Bigl[f_{\omega-\frac{\bar{\Omega}}{2}}k_i k_j
 (\ga_{\kv-\frac{\pv}{2},\omega-\frac{\bar{\Omega}}{2},\sigma}-\gr_{\kv-\frac{\pv}{2},\omega-\frac{\bar{\Omega}}{2},\sigma})\ga_{\kv+\frac{\pv}{2},\omega+\frac{\bar{\Omega}}{2},\sigma}
 \nn \\ &\qquad
 +f_{\omega+\frac{\bar{\Omega}}{2}}k_i k_j
 \gr_{\kv-\frac{\pv}{2},\omega-\frac{\bar{\Omega}}{2},\sigma}(\ga_{\kv+\frac{\pv}{2},\omega
 +\frac{\bar{\Omega}}{2},\sigma}-\gr_{\kv+\frac{\pv}{2},\omega+\frac{\bar{\Omega}}{2},\sigma})
+\delta_{ij}\frac{m}{\hbar^2}f_{\omega}(\ga_{\kv,\omega,\sigma} -\gr_{\kv,\omega,\sigma})\Bigr],
\end{align}
and 
$\gr_{\kv,\omega,\sigma}\equiv (\hbar \omega - \epsilon_{\kv,\sigma}+\frac{\rmi\hbar}{2\tau})^{-1}$ ($\sigma=\pm$ is spin index).
Expanding with respect to $\pv$ and $\bar{\Omega}$, assuming a slowly varying magnetization structure, we obtain 
\begin{align}
\xi_{ij} = &\,
\frac{\rmi e\hbar^2}{m}\sum_{\kv,\omega}\sum_\sigma \sigma \Bigl\{
f_\omega \Bigl[k_i k_j \bigl((\ga_{\kv,\omega,\sigma})^2 - (\gr_{\kv,\omega,\sigma})^2\bigr)
+\delta_{ij}\frac{m}{\hbar^2}(\ga_{\kv,\omega,\sigma} -\gr_{\kv,\omega,\sigma})\Bigr]
\nn \\&\,
+f_\omega k_i k_j \Bigl[
\frac{\hbar^2}{4m}p^2 \bigl((\ga_{\kv,\omega,\sigma})^3 - (\gr_{\kv,\omega,\sigma})^3\bigr)
+\Bigl(\frac{\hbar^2}{2m}\kv\cdot\pv\Bigr)^2 \bigl((\ga_{\kv,\omega,\sigma})^4 - (\gr_{\kv,\omega,\sigma})^4\bigr)
\Bigr]
\nn \\&\,
-\frac{\bar{\Omega}}{2}f'_\omega k_i k_j (\ga_{\kv,\omega,\sigma} - \gr_{\kv,\omega,\sigma})^2
\Bigr\}
 +O(p^3,\Omega^2).
\label{xiijeq}
\end{align}
Assuming a rotational symmetry in $k$ space and carrying out an integraton by parts with respect to $\kv$ and $\omega$, (\ref{xiijeq}) becomes  
\begin{align}
 \xi_{ij} = &\,
\frac{\rmi e\hbar^2}{m}\sum_{\kv,\omega}\sum_\sigma \sigma f'_\omega \Bigl[
-\frac{1}{12\hbar}(\delta_{ij}p^2 -p_i p_j)(\ga_{\kv,\omega,\sigma} - \gr_{\kv,\omega,\sigma})
- \frac{\bar{\Omega}}{6} \delta_{ij}k^2(\ga_{\kv,\omega,\sigma} -\gr_{\kv,\omega,\sigma})^2
\Bigr] +\Or(p^3,\Omega^2).
\end{align}
Using $\sum_{\kv}|\gr_{\kv,\sigma}|^2\simeq\frac{2\pi\nu_\sigma
\tau_\sigma}{\hbar}$ and 
$\sum_{\kv}\epsilon|\gr_{\kv,\sigma}|^4\simeq\frac{4\pi\nu_\sigma
\epsilon_{_{\rm F}\sigma}\tau_\sigma^3}{\hbar^3}$, 
($\epsilon\equiv\frac{\hbar k^2}{2m}$ and
 $\gr_{\kv,\sigma}\equiv \gr_{\kv,\omega,\sigma}|_{\omega=0}$)
 the coefficient  $\xi_{ij}$ is obtained as
\begin{align}
 \xi_{ij} 
\simeq&\;
%=&\;
 \sum_\sigma  \Bigl[
(\delta_{ij}p^2 -p_i p_j)\frac{e\hbar\nu_\sigma}{12m}
-\delta_{ij}\bar{\Omega}\frac{2\rmi e \epsilon_{_{\rm F},\sigma} \nu_\sigma
 \tau_\sigma}{3\hbar}\Bigr]
%+O(\qv^3,\bar{\Omega}^2)
,
\end{align}
where $\nu$ is 
the density of states of electron.
The result of $\jv^{12} $ is 
\begin{align}
\jv^{12} = &\,
\frac{e\hbar}{12m}\Bigl(\sum_\sigma \sigma \nu_\sigma\Bigr)
\bigl\{\nablav\times\bigl[\nablav\times\bigl(\alphav_{\R}\times\nv\bigr)\bigr]\bigr\}
-\frac{2e}{3\hbar}\Bigl(\sum_\sigma \sigma \epsilon_{_{\rm F}\sigma}
 \nu_\sigma \tau_\sigma\Bigr)\bigl(\alphav_{\R}\times\dot{\nv}\bigr).
\label{eq:j0result}
\end{align}

Next, we calculate $j^{3-6}$, (\ref{eq:j1}). 
Expanding with respect to $q$ and $p$, (\ref{eq:j1}) reduces to 
\begin{align}
j^{3-6}_i =&\,
-\frac{\rmi e\hbar^2}{m}\alpha_{\R,\ell}\epsilon_{\ell mn}\!\!
\sum_{\kv,\qv,\pv}\sum_{\omega,\Omega,\bar{\Omega}}
\rme^{-\rmi (\qv+\pv)\cdot\rv+\rmi (\Omega+\bar{\Omega}) t}
R_{no}(\pv,\bar{\Omega})
\nn \\
&\;\times f(\omega)\frac{\hbar^2}{2m}\frac{k^2}{3}
\Bigl[(q+p)_m A_{\s,i}^j(\qv,\Omega) - \delta_{im}(q+p)_p
 A_{\s,p}^j(\qv,\Omega)\Bigr]
\nn \\ &\;
\times 2{\rm Re}\tr\big[\sigma_j \ga_{\kv,\omega}\sigma_o (\ga_{\kv,\omega})^2-\sigma_o \ga_{\kv,\omega}\sigma_j
 (\ga_{\kv,\omega})^2]
\nn \\
&+O(q^3, \Omega^1). 
\label{j36eq}
\end{align}
(\ref{j36eq}) is calculated by use of the following identities,
\begin{align}
&\sum_{\kv,\omega}f(\omega)\epsilon{\rm Re}\tr
\bigl[\sigma_j \ga_{\kv,\omega}\sigma_o (\ga_{\kv,\omega})^2- \sigma_o \ga_{\kv,\omega}\sigma_j (\ga_{\kv,\omega})^2\bigr]
\nn \\
= &\;
-\frac{2}{5\hbar}\sum_{\kv,\omega}f'(\omega)\epsilon^2 {\rm Re}\tr
\bigl[\sigma_j \ga_{\kv,\omega}\sigma_o (\ga_{\kv,\omega})^2- \sigma_o \ga_{\kv,\omega}\sigma_j (\ga_{\kv,\omega})^2\bigr],
\end{align}
obtaied by an integration by parts, 
and 
\begin{align}
 2\epsilon_{joz} A_{\s,i}^j R_{no}  =&\, \nabla_i n_n,
%\end{align}
%and the next relation, 
%\begin{align}
\\
\tr[\sigma_j A \sigma_o B -\sigma_o A \sigma_j B]
=&\;
2i\epsilon_{joz}\sum_\sigma \sigma A_{-\sigma}B_\sigma,
\end{align}
where $A\equiv {\rm diag}(A_+,A_-)$ and $B\equiv {\rm diag}(B_+,B_-)$
are any 2$\times$2 diagonal matrices.
The leading contribution of (\ref{eq:j1}) then reads 
\begin{align}
 \jv^{3-6} \!=&\,
\eta \nablav\times\bigl[\nablav\times\bigl(\alphav_{\R}\times\nv\bigr)\bigr],
\label{eq:j1result}
\end{align}
where the coefficient is %given as 
\begin{align}
\eta =& \frac{2e\hbar}{15\pi m}\sum_{\kv}
{\rm Im}\Bigl[\sum_\sigma \sigma \epsilon^2 \ga_{\kv,-\sigma}(\ga_{\kv,\sigma})^2\Bigr]
\nn \\
=& \;
\frac{e\hbar}{60m\Delta_{\rm sd}^2}\sum_\sigma \sigma
\Bigl(\epsilon_{_{\rm F},\sigma}^2 \nu_{\sigma}
-5\epsilon_{_{\rm F},\sigma}\epsilon_{_{\rm F},-\sigma} \nu_{\sigma}
\Bigr),
\label{eq:eta}
\end{align}
and we used 
$\sum_{\kv}\epsilon^2 \ga_{\kv,-\sigma} (\ga_{\kv,\sigma})^2 \simeq 
\frac{\pi \rmi}{8\Delta_{\rm sd}^2}(2\nu_{-\sigma}\epsilon_{_{\rm F},-\sigma}^2
+3\epsilon_{_{\rm F},\sigma}^2 \nu_{\sigma}
-5\epsilon_{_{\rm F},\sigma}\epsilon_{_{\rm F},-\sigma} \nu_{\sigma}
)$.

Therefore %we obtained the electric current as
the total local electric current density is
\begin{align}
 \jv =&
 (\xi_1+\eta)\bigl\{\nablav\times\bigl[\nablav\times\bigl(\alphav_{\R}\times\nv\bigr)\bigr]\bigr\}
 +\xi_2 \bigl(\alphav_{\R}\times\dot{\nv}\bigr)
%+\jv^{\rm diff}
,
\label{eq:ecurrent_by_Rashba}
\end{align}
where $\xi_1 = \frac{e\hbar}{12m}\Bigl(\sum_\sigma \sigma \nu_\sigma\Bigr)$,
$\xi_2 = -\frac{2e}{3\hbar}\Bigl(\sum_\sigma \sigma \epsilon_{_{\rm F}\sigma}
 \nu_\sigma \tau_\sigma\Bigr)$.
% and $\eta$ is given by Eq.(\ref{eq:eta}).

Electric current driven by effective electromagnetic fields is generally  written in diffusive regime as follows:
\begin{align}
\jv =&\;
\sigma_{{\rm s}}\Ev_{\rm s} 
+\frac{1}{\mu_{\rm s}} \nablav\times\Bv_{\rm s}
- D_{\rm s} \nablav \rho_{{\rm s}},
\label{eq:jbyU1}
\end{align}
where $\Ev_{\rm s} $ and $\Bv_{\rm s} $ represent driving fields,
 $\sigma_{\rm s}$ is the conductivity for spin, $\mu_{\rm s}$ is the
magnetic permeability of spin magnetic field
and the last term is a diffusive contribution 
($D_{\rm s}$ is the diffusion constant for spin and $\rho_{\rm s}$
is the density of electron spin). 
Comparing our result, ({\ref{eq:ecurrent_by_Rashba}), to (\ref{eq:jbyU1}), we see that 
\begin{align}
\sigma_{\rm s} \Ev_{{\rm s}\pm} &=
\xi_2 \bigl(\alphav_{\R}\times\dot{\nv}\bigr) \nonumber\\
\frac{\Bv_{\rm s} }{\mu_{\rm s}} 
&= (\xi_1+\eta)\bigl[\nablav\times\bigl(\alphav_{\R}\times\nv\bigr)\bigr].
\label{EsBsres0}
\end{align}
We know that spin conductivity is given by
\begin{align}
\sigma_{\rm s} =& \sum_{\sigma=\pm}\sigma\frac{e^2\hbar^2}{3m^2}k_{{\rm
F}\sigma}^2\nu_\sigma\tau_\sigma,
\label{eq:sigmaB}
\end{align}
and thus the spin electric field reads
\begin{align}
\Ev_{{\rm s}} &= - \frac{m}{e\hbar}\alphav_{\rm R}\times\dot{\nv}.
\label{Esresult}
\end{align}
This result agrees with result of direct estimate of spin motive force \cite{Kim12,Tatara_smf13}.
In contrast, to identify spin magnetic field from  (\ref{EsBsres0}), 
we need additional information on the permeability.
This is accomplished by analyzing the spin Hall effect, which is carried out in the next section.

%%%%%%%%%%%%%%%%%%%%%%%%%%%%%%%%%%%%%%%%%%%%%%%%%%%%%%%%%%%%%%%%%%%%%%%%%
\section{The spin Hall effect induced by spin magnetic filed} \label{sec:hallcurrent}
To determine the effective spin magnetic field, (\ref{EsBsres0}) is not sufficient.
In this section we calculate the Hall effect caused by the spin magnetic field when an electric field, $\Ev$, is applied and determine the spin magnetic field uniquely.
Since the Hall effect studied here drives electron spin, the effect is spin Hall effect.  
The Lagrangian has now the following additional terms coming from an applied vector potential, $\Av$ 
($\Ev=-\,\,\dot{\!\!\Av}$),
\begin{align}
\delta{\cal L} = & \int d^3 r 
\biggl[\frac{\rmi e\hbar}{2m}A_j\bigl(\adag\nablalr_{\! j} a\bigr)
-\frac{e^2}{2m}\bigl(\adag A^2 a\bigr)
-\frac{e\hbar}{m}\bigl(\adag A_j A_{\s,j}a\bigr)
+\frac{e}{\hbar}\alpha_{_{\rm R},j}
\epsilon_{jkl}R_{ln}A_k\bigl(\adag\sigma_n a\bigr)
\biggr].
\end{align}
The electric current density is also modified to be $j_i+\delta j_i$, where 
\begin{align}
\delta j_i =&
-\frac{e^2}{m}A_i\average{\adag a}.
\end{align}
We calculate the Hall current induced by the effective magnetic field
with applied electric field which is spatially homogeneous.
%%%%%%%%%%%%%%%%%%%%%%%%%%%%%%%%%%%%%%%%%%%%%%%%%%%%%%%
\begin{figure}[hbp]
%  \psfrag{R}{$\alpha_{\R}$}
\begin{center}
 \includegraphics[width=160pt]{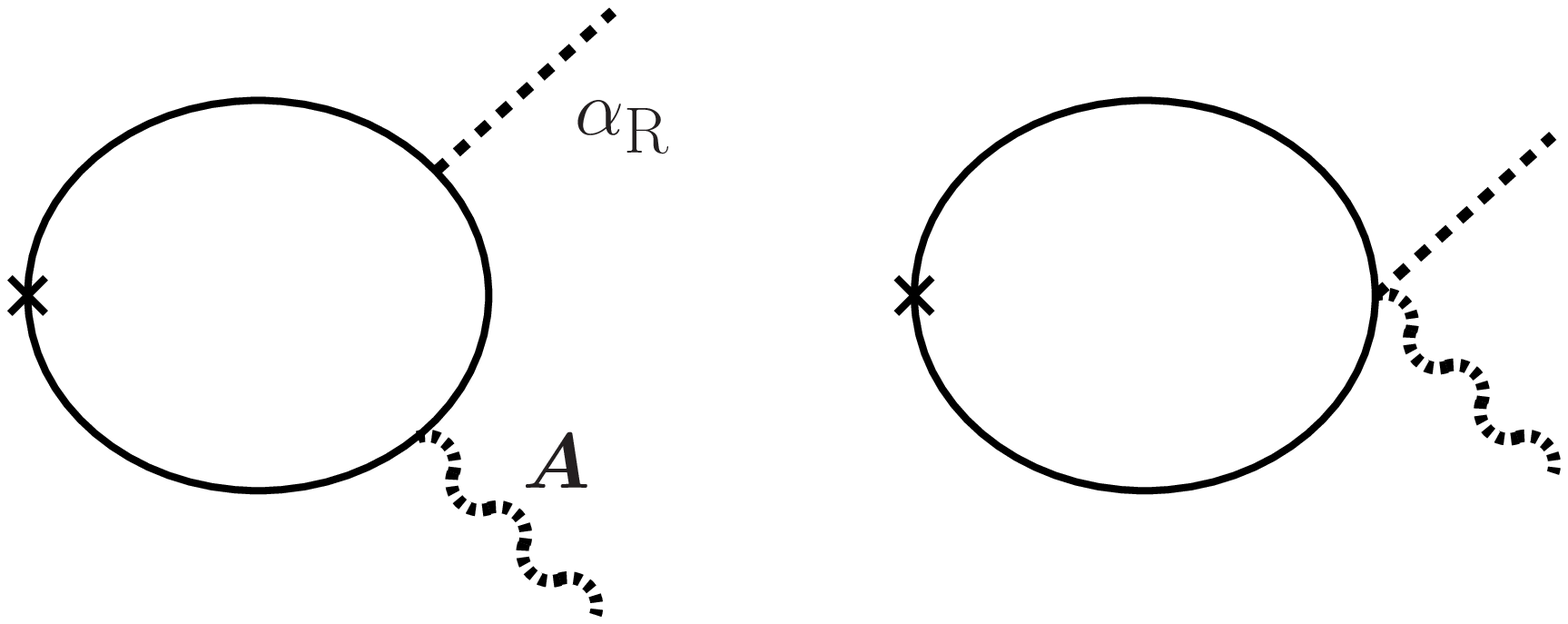}
 \includegraphics[width=160pt]{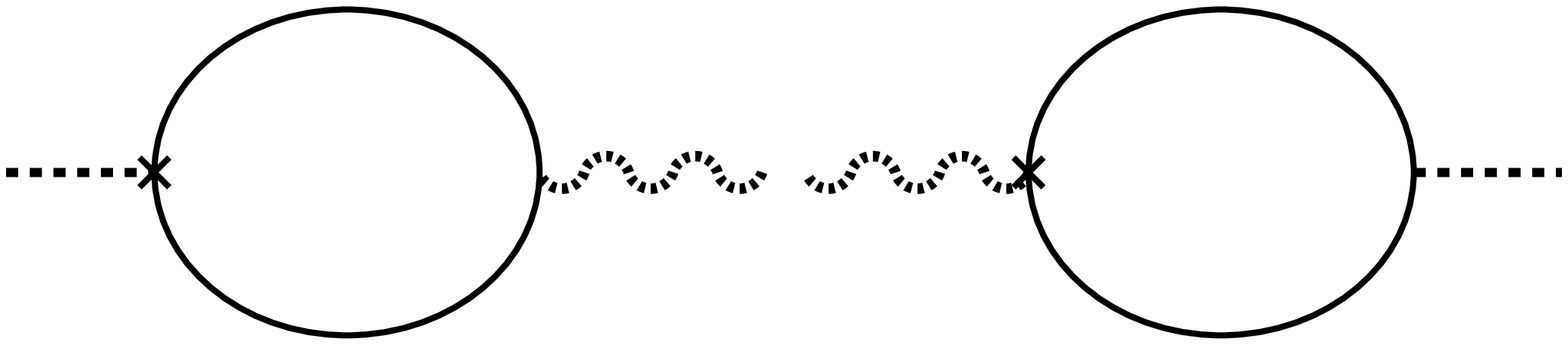}
  \caption{The Feynman diagrams representing the leading contributions to the spin Hall 
current. Dotted line denotes the Rashba interaction, and dotted wave line represents applied electric field, written by use of a vector potential, $A$.
Contributions containing the interaction with spin gauge field, $A_{\rm S}$, are neglected since they turn out to be smaller by the order of $\hbar/(\ef \tau)$.
\label{fig:halljlocal_dia_main}}
\end{center}
\end{figure}
%%%%%%%%%%%%%%%%%%%%%%%%%%%%%%%%%%%%%%%%%%%%%%%%%%%%%%%
The leading contribution  to the Hall current density, $j_i^{{\rm
hall}}$, described by the Feynmann diagrams in figure \ref{fig:halljlocal_dia_main}, reads
\begin{align}
j_i^{{\rm hall}} &=
-\frac{\rmi e^2\hbar}{m}\sum_{\kv,\pv,\omega,\Omega}
\rme^{-\rmi\pv\cdot\rv+\rmi\Omega t}\alpha_{{\rm R},j}\epsilon_{jkl} \nn \\
&\times\tr\biggl\{
\frac{\hbar^2}{m}k_i k_k A_m(\Omega)R_{ln}(\pv)
\biggl[
\biggl(k-\frac{p}{2}\biggr)_m
 g_{\kv-\frac{\pv}{2},\omega-\frac{\Omega}{2}}g_{\kv-\frac{\pv}{2},\omega+\frac{\Omega}{2}}\sigma_n
 g_{\kv+\frac{\pv}{2},\omega+\frac{\Omega}{2}}
\nn \\ & \qquad\qquad
+\biggl(k+\frac{p}{2}\biggr)_m g_{\kv-\frac{\pv}{2},\omega-\frac{\Omega}{2}}\sigma_n g_{\kv+\frac{\pv}{2},\omega-\frac{\Omega}{2}}g_{\kv+\frac{\pv}{2},\omega+\frac{\Omega}{2}}
\biggr]
\nn \\
&\qquad\quad -k_i A_k(\Omega)R_{ln}(\pv) g_{\kv-\frac{\pv}{2},\omega-\frac{\Omega}{2}}\sigma_n g_{\kv+\frac{\pv}{2},\omega+\frac{\Omega}{2}}
+\delta_{ik}k_m A_m(\Omega)R_{ln}(\pv) \sigma_n g_{\kv,\omega-\frac{\Omega}{2}}
 g_{\kv,\omega+\frac{\Omega}{2}}
\nn \\
&\qquad\quad-k_k A_i(\Omega)R_{ln}(\pv) g_{\kv-\frac{\pv}{2},\omega}\sigma_n g_{\kv+\frac{\pv}{2},\omega}
\biggr\}^<
,
\label{eq:hall_current_As_0}
\end{align}
The contributions containing spin gauge field at the linear order are neglected, because those are smaller than the ones in (\ref{eq:hall_current_As_0}) by the order of
$\frac{\hbar}{\epsilon_{_{F}}\tau}$.
Expanding with respect to external wave vector and frequency, $p$ and $\Omega$, we obtain
\begin{align}
j_i^{{\rm hall}} &=
-\frac{\rmi e^2\hbar}{m}\sum_{\kv,\pv,\omega,\Omega}
\rme^{-\rmi\pv\cdot\rv+\rmi\Omega t}\alpha_{{\rm R},j}\epsilon_{jkl} \nn \\
&\times\tr\biggl\{
\frac{\hbar^2}{2m}k_i k_k p_m A_m(\Omega)R_{ln}(\pv)
\bigl[
- g_{\kv-\frac{\pv}{2},\omega-\frac{\Omega}{2}}g_{\kv-\frac{\pv}{2},\omega+\frac{\Omega}{2}}\sigma_n g_{\kv+\frac{\pv}{2},\omega+\frac{\Omega}{2}}
+ g_{\kv-\frac{\pv}{2},\omega-\frac{\Omega}{2}}\sigma_n g_{\kv+\frac{\pv}{2},\omega-\frac{\Omega}{2}}g_{\kv+\frac{\pv}{2},\omega+\frac{\Omega}{2}}
\bigr]
\nn \\
&\qquad\quad -k_i (\kv\cdot\pv) A_k(\Omega)R_{ln}(\pv) 
g_{\kv-\frac{\pv}{2},\omega-\frac{\Omega}{2}}\sigma_n g_{\kv+\frac{\pv}{2},\omega+\frac{\Omega}{2}}/(\kv\cdot\pv)
\biggr\}^<
+\Or(\Omega^2,p^2).
\end{align}
The lesser components of (\ref{eq:hall_current_As_0}) are calculated as
\begin{align}
&[g_{\kv-\frac{\pv}{2},\omega-\frac{\Omega}{2}}g_{\kv-\frac{\pv}{2},\omega+\frac{\Omega}{2}}\sigma_n
 g_{\kv+\frac{\pv}{2},\omega+\frac{\Omega}{2}}]^<
\nn \\ =&
(f_{\omega+\frac{\Omega}{2}}-f_{\omega-\frac{\Omega}{2}})
\gr_{\kv-\frac{\pv}{2},\omega-\frac{\Omega}{2}}\ga_{\kv-\frac{\pv}{2},\omega+\frac{\Omega}{2}}\sigma_n
 \ga_{\kv+\frac{\pv}{2},\omega+\frac{\Omega}{2}}
\nn \\ &
+f_{\omega-\frac{\Omega}{2}}\ga_{\kv-\frac{\pv}{2},\omega-\frac{\Omega}{2}}\ga_{\kv-\frac{\pv}{2},\omega+\frac{\Omega}{2}}\sigma_n \ga_{\kv+\frac{\pv}{2},\omega+\frac{\Omega}{2}}
+f_{\omega+\frac{\Omega}{2}}\gr_{\kv-\frac{\pv}{2},\omega-\frac{\Omega}{2}}\gr_{\kv-\frac{\pv}{2},\omega+\frac{\Omega}{2}}\sigma_n \gr_{\kv+\frac{\pv}{2},\omega+\frac{\Omega}{2}},
\\
&[g_{\kv-\frac{\pv}{2},\omega-\frac{\Omega}{2}}\sigma_n
 g_{\kv+\frac{\pv}{2},\omega+\frac{\Omega}{2}}]^<
\nn \\ =&
(f_{\omega+\frac{\Omega}{2}}-f_{\omega-\frac{\Omega}{2}})
\gr_{\kv-\frac{\pv}{2},\omega-\frac{\Omega}{2}}\sigma_n
 \ga_{\kv+\frac{\pv}{2},\omega+\frac{\Omega}{2}}
\nn \\ &
+f_{\omega-\frac{\Omega}{2}}\ga_{\kv-\frac{\pv}{2},\omega-\frac{\Omega}{2}}\sigma_n \ga_{\kv+\frac{\pv}{2},\omega+\frac{\Omega}{2}}
+f_{\omega+\frac{\Omega}{2}}\gr_{\kv-\frac{\pv}{2},\omega-\frac{\Omega}{2}}\sigma_n \gr_{\kv+\frac{\pv}{2},\omega+\frac{\Omega}{2}}.
\end{align}
Assuming a rotational symmetry 
(i.e., using $k_i k_j =\frac{k^2}{3}\delta_{ij}$), 
Hall current density reduces to 
\begin{align}
j_i^{{\rm hall}} &=
-\frac{\rmi e^2\hbar}{3m}\sum_{\kv,\pv,\Omega}
\rme^{-\rmi\pv\cdot\rv+\rmi\Omega t}\Omega \epsilon\Gamma_1
\bigl\{\Av(\Omega)\times\bigl[\pv\times
(\alphav_{\R}\times\nv(\pv))\bigr]\bigr\}_i
,
\end{align}
where 
\begin{align}
\Gamma_1 &\equiv \sum_{\omega,\sigma} \sigma\bigl[
\hbar f_\omega \bigl((\ga_{\kv,\omega,\sigma})^4-(\gr_{\kv,\omega,\sigma})^4\bigr)-f'_\omega\bigl(\gr_{\kv,\omega,\sigma}
 (\ga_{\kv,\omega,\sigma})^2-(\gr_{\kv,\omega,\sigma})^2 \ga_{\kv,\omega,\sigma}\bigr)\bigr].
\end{align}
The summation about $\omega$ and $k$ is evaluated as follows;
\begin{align}
 \sum_{\kv}\epsilon\Gamma_1
&=\frac{1}{3}\sum_{\kv,\omega,\sigma}\sigma
f'_\omega \epsilon(\ga_{\kv,\omega,\sigma} - \gr_{\kv,\omega,\sigma})^3
\nn \\
&\simeq
\frac{2i}{\hbar^2}\sum_\sigma \sigma 
 \epsilon_{_{\rm F}\sigma}\nu_\sigma \tau^2_\sigma.
\end{align}
The Hall current is finally obtained as 
\begin{align}
\jv^{{\rm hall}} & =
-\sum_\pm
 (\pm)\frac{e\tau_\pm}{m}\sigma_{_{\B}\pm}\frac{m}{e\hbar}\bigl(\Ev\times(\nablav\times(\alphav\times\nv))\bigr),
\label{eq:hall_effect_result}
\end{align}
where $\sigma_{_{\B}\sigma}\equiv\frac{2e^2}{3m}\epsilon_{_{\rm
F}\sigma}\nu_\sigma \tau_\sigma$ is the spin-resolved Boltzmann conductivity.
Thus the Hall effect is described by a standard expression of
\begin{align}
\jv^{{\rm hall}} =\sigma_{\rm H}(\Ev\times\Bv_{\rm s}),
\label{halleffect}
\end{align}
 where 
$\sigma_{\rm H}\equiv - \sum_\pm
 (\pm)\frac{e\tau_\pm}{m}\sigma_{_{\B}\pm}$
and
\begin{align}
\Bv_{\s}\equiv  \frac{m}{e\hbar}\nablav\times(\alphav_{\R}\times\nv).
\label{Bsresult}
\end{align}
In terms of Hall electric field,
(\ref{eq:hall_effect_result}) 
is written as 
\begin{align}
% E^{\rm hall} = -\frac{1}{ne}B_{\s}j, 
 \Ev^{\rm hall} = -\frac{1}{ne}\jv\times\Bv_{\s}, 
\label{eq:normalHalleffect}
\end{align}
where $n$ is electron density,  
$j$ is the magnitude of longitudinal 
electric current driven by applied electric field.
(\ref{halleffect}) and (\ref{eq:normalHalleffect}) are the central results of the present paper, indicating that the field $\Bv_{\rm s}$ exerts the Lorentz force of $\Fv=-e\vv\times \Bv_{\rm s}$ for electron spin and that $\Bv_{\s}$ certainly
play a role as a magnetic field for conduction electron.

%-----------------------------------------------------
\section{Rashba-induced effective gauge field}
 \label{sec:fromgauge}
In the preceeding sections, 
we have demonstrated that the effective spin electromagnetic fields 
are determined by calculating transport properties.
Our results, (\ref{Esresult}) and (\ref{Bsresult}), indicate that the Rashba-induced effective field is written as
\begin{align}
\Esv &= -\dot{\Av}_{\rm R} \nonumber\\
\Bsv &= \nablav \times \Av_{\rm R},
\label{Aexpression}
\end{align}
with 
\begin{align}
\Av_{\rm R} &\equiv  \frac{m}{e\hbar}(\alphav_{\R}\times\nv).
\label{ARdef}
\end{align}
The present spin electromagnetic fields are therefore explained by a standard U(1) gauge theory, as was argued in \cite{Kim12}.
This fact is not obvious, since the combination of the Rashba spin-orbit interaction and sd interaction does not necessarily lead to an emergence of U(1) gauge symmetry.
Neverthelss a gauge field scenario holds as far as non-linear effects of the Rashba interaction are neglected.
In fact, the Rashba interaction has the same effect as an SU(2) gauge field if non-linear effects are neglected. 
This is easily seen from (\ref{eq:Lgaugetrans}), which indicates that the Rashba interaction in a rotated frame is written as
\begin{align}
\int d^3 r \frac{\rmi\hbar^2}{2m}\Av_{\rm R}^{\ell}\cdot (a^\dagger\sigma_{\ell} \bnablalr a),
\end{align}
where 
$A_{\R,j}\equiv A_{\R,j}^{\ell}\sigma_{\ell} \equiv
        -\frac{m}{\;\hbar^2}\alpha_{\R,i}\epsilon_{ijk}R_{k\ell}\sigma_{\ell}$.
The kinetic term of  (\ref{eq:Lgaugetrans}) thus is written by an SU(2) gauge field defined as $\tilde{\Av}_{\rm R}\equiv \Asv+\Av_{\rm R}$ as 
\begin{align}
\int d^3 r \frac{\hbar^2}{2m}a^\dagger  (\nablav+\rmi\tilde{\Av}_{\rm R})^2 a +\Or((\Av_{\rm R})^2) .
\end{align}
In the strong sd coupling limit, therefore, $\tilde{\Av}_{\rm R}^z$ acts as an effective U(1) gauge field and effective electromagnetic fields emerge according to 
$\tilde{\Ev}_{\rm R}=-\dot{\tilde{\Av}}_{\rm R}^z=\Ev_{\rm B}+\Esv$ and 
$\tilde{\Bv}_{\rm R}=\nablav\times \tilde{\Av}_{\rm R}^z=\Bv_{\rm B}+\Bsv$, where  
$\Esv$ and $\Bsv$ are the Rashba contributions (\ref{Aexpression}),  
$\Ev_{\rm B}\equiv -\dot{\Av}_{\rm s}^z$ and 
$\Bv_{\rm B}\equiv \nablav \times {\Av}_{\rm s}^z$ are the Berry's phase contributions discussed by Volovik \cite{Volovik87}.
It is generally expected that a novel U(1) gauge field different from
the adiabatic gauge field emerges when spin-orbit interaction is
included, as long as the linear effects are concerned.
%%%

\section{Experimental possibilities}

The Rashba-induced spin electromagnetic field we have discussed is an extension of the spin Berry's phase generalized to include the Rashba interaction.
The fields is linear in the gradient in space or in time and thus becomes dominant in slowly varying magnetization structures, since conventional spin Berry's phase is second order of gradients.
Let us estimate the magnitude. Rashba interaction is large on the surface of heavy metals in particular when doped with Bi \cite{Ast07}.
Choosing $\alpha_{\rm R}=3$ eV\AA \cite{Kim12}, we obtain 
$\frac{m\alpha_{\rm R}}{e\hbar}=2.5\times 10^{-6}$ Vs/m.
For a magnetization dynamics with angular frequency $\omega$ of 1 GHz, we thus expect a large spin electric field of $E_{\rm S}=\frac{m\alpha_{\rm R}}{e\hbar}\omega=2.5$ kV/m.
For a magnetization structure having a typical length scale $\lambda$ of 1nm, the spin magnetic field corresponds to an extremely high field of $B_{\rm S}=\frac{m\alpha_{\rm R}}{e\hbar}\lambda=2.5$ kT.
The Rashba interaction is thus useful in creating extremely high effective spin electromagnetic fields.

When a magnetic structure is flows, the spin electric field is induced.
In the case of a flow without deformation, the magnetization vector depends on the time as $\nv(\rv,t)=\nv(\rv-\vv t)$, where $\vv$ is velocity of the flow.
The spin electric field then reads
\begin{align}
\Ev_{\rm s} &= \frac{m}{e\hbar}(\alphav_{\R}\times(\vv\cdot\nabla)\nv).
\label{Eflow}
\end{align}

\subsection{Vortices}

A crucial difference between the conventional spin Berry's phase and Rashba-induced one is that the former is of topological origin while the latter is not.
The Rashba-induced field is therefore expected to be useful to discriminate topologically equivalent magnetic structures. 
An example is magnetic vortex or skyrmion.
Structures shown in figure \ref{FIGvortex}(a) and (b) are topologically equivalent, since structure (b) is obtained by shifting the in-plane angle $\phi$ of magnetization by $\frac{\pi}{2}$.
They have, however, different response to the Rashba-induced field. 
Choosing $\alphav$ perpendicular to the plane, the Rashba-induced spin magnetic field measures the divergence of the magnetization structure resulting in a vanishing for structure (a) while that of (b) is finite. 
The total flux for a vortex (b) is $\Phi\equiv\int_S d\Sv\cdot\Bsv=\int_Cd\rv\cdot\Asv$ where $S$ is the area of the vortex and $C$ is its perimeter.
Using (\ref{ARdef}), the flux is 
$\Phi=\frac{m}{e\hbar}\alpha_{\rm R} L=\frac{\hbar}{e}\frac{\alpha_{\rm R} \kf^2 L}{2\ef}$, where $L$ is the length of the perimeter of a vortex.

When a vortex flows, the Rashba-induced spin electric field,  (\ref{Eflow}), vanishes, since the average of $\nabla \nv$ inside a vortex vanishes if not deformed.
This feature is a significant difference from the conventional Berry's phase contribution (topological spin motive force), which has been used to detect skyrmion motions \cite{Schulz12}.

%%%%%%%%%%%%%%%%%%%%%%%%%%%%%%%%%%%%%%%%%%%%%%%%%%%%%%%
\begin{figure}[htbp]
\begin{center}
  \includegraphics[width=0.5\hsize]{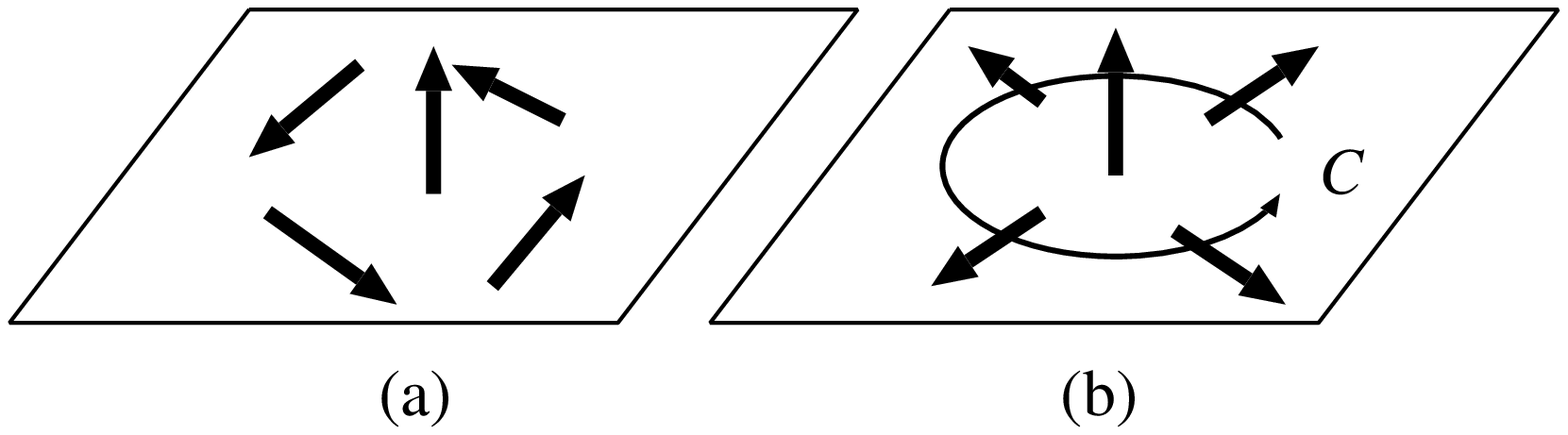}  
\caption{ Schematic picture of two structures of vortex which are topologically equivalent, i.e., having equal spin Berry's phase. The magnetization at the center is pointing perpendicular to the plane.
The Rashba-induced spin magnetic field vanishes for a structure (a), while it is finite for a structure (b), where the total flux of spin magnetic field is $\Phi=\frac{m}{e\hbar}\alpha_{\rm R} L$ ($L$ is perimeter of a vortex). 
\label{FIGvortex}} 
 \includegraphics[width=0.5\hsize]{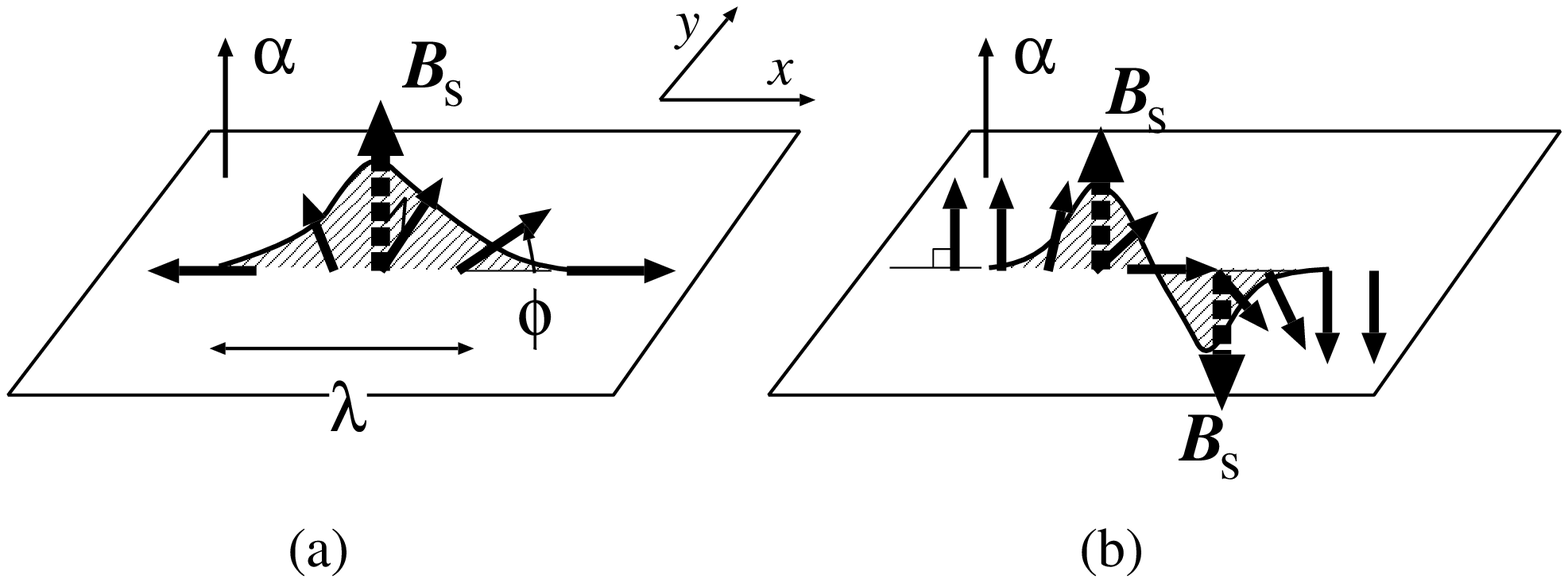} 
 \caption{ 
(a) Domain walls with in-plane easy axis  and (b) perpendicular easy axis (out-of-plane Neel wall). 
The spin magnetic field is induced inside the wall proportional to $\frac{\partial n_x}{\partial x}$, where $n_x$ is $x$ component of magnetization.
Curves are schematically shown profiles of spin magnetic field induced by the wall. 
\label{FIGdw}}
\end{center}
\end{figure}

\subsection{Domain walls}

Let us discuss the spin magnetic field for domain wall in a film in the $xy$ plane with Rashba field in the $z$ direction. 
The magnetization of the domain wall is changing in the $x$ direction but is uniform in the $y$ direction.
We first consider an in-plane domain wall as in figure \ref{FIGdw}(a).
The domain wall profile is $\theta=\frac{\pi}{2}$, 
$\cos\phi=\tanh \frac{x}{\lambda}$,
where $\lambda$ is thickness of the wall.
The spin magnetic field then reads ($n_x=\sin\theta\cos\phi$)
$\Bsv=\frac{m\alpha_{\rm R}}{e\hbar \lambda }\frac{\partial n_x}{\partial x}\hat{\boldsymbol{z}}=
\frac{m\alpha_{\rm R}}{e\hbar \lambda }\frac{1}{\cosh^2 \frac{x}{\lambda}}\hat{\boldsymbol{z}}$ ($\hat{\boldsymbol{z}}$ is unit vector in the $z$ direction).
The spin magnetic field is therefore localized to the wall, and it corresponds to a high field of 250 T if $\lambda=10$nm.
This localized strong field would be detected as a local spin Hall voltage in the $y$ direction when current is applied in the $x$ direction.

When the in-plane domain wall flows in $x$ direction with a speed of $v_x$, a spin voltage in $y$ direction (defined as $V_y\equiv \int dx E_{{\rm s},y}$ with  (\ref{Eflow})) is induced;
\begin{align}
V_y=\frac{2m\alpha_{\rm R}}{e\hbar}v_x.
\end{align}
For $\alpha_{\rm R}=3$ eV\AA and $v_x=100$ m/s, the voltage is $0.5$ mV. 
This value is 1000 times larger than the conventional Berry's phase contribution observed in a permalloy, $0.4\mu$V at 130m/s.
Even for a system having a moderate value of $\alpha_{\rm R}=3$ meV\AA, therefore, the Rashba-induced signal is comparable to the conventional signal. 

In the case of out-of-plane domain wall of Neel type (figure \ref{FIGdw}(b)), 
 $\cos\theta=\tanh \frac{x}{\lambda}$ with magnetization at the center of the wall pointing $x$ direction, 
we have
$\Bsv= - \hat{\boldsymbol{z}}\frac{m\alpha_{\rm R}}{e\hbar \lambda }
 \frac{\sinh \frac{x}{\lambda}}{\cosh^2 \frac{x}{\lambda}} $.
The field produced by an out-of-plane wall changes sign  at the wall center and has a large field gradient.
For a flowing out-of-plane wall, the spin voltage, (\ref{Eflow}),  vanishes.

\subsection{Effect of gradient of Rashba field}
So far we have argued the spin magnetic field induced by magnetization structures. 
(\ref{Aexpression}) inducates that it arises also when the Rashba field, $\alphav_{\rm R}$, has a gradient, which is naturally expected at the surfaces of thin film \cite{Henk03,Bihlmayer06,Kosugi11}.
The spin magnetic field then becomes
\begin{align}
\Bsv=\frac{m}{e\hbar}((\nv\cdot\nablav)\alphav_{\rm R}-\nv(\nablav\cdot\alphav_{\rm R}))
\end{align}
We consider a thin film in the $xy$ plane with $\alphav_{\rm R}=(0,0,\alpha_{\rm R}(z))$ along the $z$ axis. When $\nv$ is within the $xy$ plane, we obtain 
$\Bsv=-\frac{m}{e\hbar}(\nabla_z \alphav_{\rm R})\nv$.
If the Rashba interaction decays at the length $d$, we might approximate $\nabla_z \alphav_{\rm R}\sim 2\alpha_{\rm R}/d$, resulting in a magnetic field of 25kT if $d=0.1$ nm.
%Such a high field is expected to induce extremely large spin Hall effect as was recently predicted in Dirac fermion system of Bi \cite{Fuseya12}.

\section{Propagation of spin electromagnetic field}

According to our result,  (\ref{Aexpression}), spin electromagnetic fields satisfy a conventional Faraday's law, 
$\nablav\times \Esv+\frac{\partial \Bsv}{\partial t}=0$.
There is therefore no monopole in the present system.

From our results, (\ref{EsBsres0}) and (\ref{Bsresult}), we see that
\begin{align}
\frac{1}{\mu_{\s}} =  \frac{e\hbar}{m}(\xi_1+\eta). 
\label{musresult}
\end{align}
Electric permittivity in a diffusive case is  $\epsilon_{\rm s}=\sigma_{\rm s}\tau$.
The speed of the spin electromagnetic field is defined as
\begin{align}
c_{\rm s}\equiv \frac{1}{\sqrt{\epsilon_{\rm s}\mu_{\rm s}}}.
\end{align}
It should be noted that the sign of $\epsilon_{\rm s}$ and $\mu_{\rm s}$ may be negative depending on the material. 
If the product of the two is positive, the spin electromagnetic wave propagates, while it does not if one of the two is negative. 

Let us look into an example of a strong sd coupling limit, $\nu_-=0$ (i.e., $\Delta_{{\rm sd}}=\ef$). 
In this limit, 
$\frac{1}{\mu_{\s}} =  \frac{e^2\hbar^2 }{60m^2}\nu_+
\lt(\frac{\epsilon_{{\rm F}+}^2}{\Delta_{\rm sd}^2}+5\rt) $.
In the present model of a parabolic band, $\epsilon_{{\rm F}+}/\Delta_{\rm sd}=2$.
Approximating
$\frac{1}{\mu_{\s}} \simeq  \frac{3e^2\hbar^2}{20m^2}\nu  $  and $\epsilon_{\rm s}\simeq \frac{e^2\hbar^2}{3m^2}\kf^2\nu \tau^2$
($\nu$, $\kf$ and $\tau$ here are spin-averaged quantities),
we obtain 
\begin{align}
c_{\s} = \frac{3}{2\sqrt{5}} 
\frac{1}{\kf\tau}.
\label{cs}
\end{align}
For $\kf^{-1}=1.5 \AA$, and $\tau=10^{-15}$ s, 
the spin electromagnetic wave propagates with a speed of $c_{\s}=1\times10^5$ m/s.
This speed is larger than in the weak coupling regime discussed in  \cite{Tatara12}.
Observation of the propagation speed of spin electric field is an interesting challenge for experimentalists.

\section{Summary}

We have derived an expression for spin electromagnetic field induced by the Rashba spin-orbit interaction and magnetization structures in the strong sd coupling limit by calculating the pumped current and the spin Hall effect.
Spin relaxation is not considered.
We found that the spin electromagnetic fields are described totally by an effective U(1) gauge field, $\Av_{\rm R} =  \frac{m}{e\hbar}(\alphav_{\R}\times\nv)$ at the linear order in the Rashba interaction.
Thus a naive picture of regarding the Rashba interaction as an effective gauge interaction is valid at the linear order.
The spin electromagnetic field discussed here is a generalized spin Berry's phase.
In contrast to the spin Berry's phase, the Rashba-induced one is linear in the gradient of the magnetization profile, and can generate extremely high electric and magnetic fields of kV/m and kT for a frequency of 1GHz and for a structure of nano meter scale, respectively. 

Since the spin electromagnetic fields have U(1) gauge invariance, there is no monopole in the strong sd limit without spin relaxation, in contrast to what was observed in  \cite{Takeuchi12} in the weak sd coupling regime.
The present result confirm the argument in  \cite{Takeuchi12} that spin relaxation is essential for an emergence of monopole.
In fact, the spin electric field in the presence of spin relaxation,
$\Esv\propto \alphav_{\rm R}\times(\nv\times\dot{\nv})$, cannot be written by a time derivative of any local quantity, resulting in non-vanishing 
$\nablav\times\Esv+\dot{\Bsv} \equiv -\jv_{\rm m}$ for any local function $\Bsv$.
To study the effect of spin relaxation in the present framework is an urgent and important future work.

\section*{Acknowledgment}
The authors thank H. Ueda, H. Kohno and H. Kino for valuable discussion. 
N. N. is financially supported by the Japan Society for the Promotion of Science for Young Scientists.
This work was supported by 
 a Grant-in-Aid for Scientific Research (C) (Grant No. 25400344) from Japan Society for the Promotion of Science and UK-Japanese Collaboration on Current-Driven Domain Wall Dynamics from JST.

%%%%%%%%%%%%%%%%%%%%%%%%

\section*{References}

\bibliographystyle{unsrt}

\end{document}